\documentclass[aps,pra,floatfix,twocolumn,nofootinbib]{revtex4-1}

\usepackage{graphicx}
\usepackage{amssymb,amsmath,amsfonts,mathrsfs,fancyhdr}
\usepackage{braket}

\usepackage{epsf}
\usepackage{color}

\def\sH{\mathscr{H}}

\def\Tr{\operatorname{Tr}}

\definecolor{darkgreen}{RGB}{0, 150, 0}

\begin{document}
\title{A counterintuitive role of geometry in transport by quantum walks}
\author{J. Mare\v s, J. Novotn\'y, M. \v Stefa\v n\' ak, I. Jex}
\affiliation{Department of Physics, Faculty of Nuclear Sciences and Physical Engineering, Czech Technical University in Prague, B\v rehov\'a 7, 115 19 Praha 1 - Star\'e M\v esto, Czech Republic}
\date{\today}

\begin{abstract}
Quantum walks are accepted as a generic model for quantum transport. The character of the transport crucially depends on the properties of the walk like its geometry and the driving coin. We demonstrate that increasing transport distance between source and target or adding redundant branches to the actual graph may surprisingly result in a significant enhancement of transport efficiency. We explain analytically the observed non-classical effects using the concept of trapped states for several intriguing geometries including the ladder graph, the Cayley tree and its modifications.
\end{abstract}

\maketitle

Transport of matter, energy or information are key ingredients involved at various stages of fundamental natural processes like photosynthesis \cite{transport_biology, Croce2018}, designed engines \cite{Plawsky}, communication channels and information processing protocols \cite{Gupta,Chuang}. Its characteristic properties are established by the given underlying transport mechanism. In the quantum domain, the superposition principle leads to mutual interference of available trajectories of the transported microscopic objects resulting in rich 
transport behavior \cite{Nazarov}. An important role in understanding the transport mechanism is assigned to so-called trapped states (also called invariant localised or dark states) \cite{Inui2004,Caruso2009,Poltl2009,Huelga,Creatore2013,Mendoza2013}, i.e. eigenstates of a particular quantum evolution, whose support is not spread over the whole medium. Due to their existence, the efficiency of the quantum transport may be significantly reduced \cite{Rebentrost2009,Chin2010}. 

In this paper we explore the role of the medium's geometry in the formation of trapped states and the overall source-to-sink excitation transport efficiency (further in the text simply transport efficiency). The main aim is to present two interesting effects lacking classical counterpart. We show that extension of the transporting medium or adjunction of, for example, redundant branches of the graph representing the medium may surprisingly increase transport efficiency. As far as we know these effects have not been reported yet. We investigate these transport phenomena in the context of coined (CQWs) \cite{review} and percolated coined quantum walks (PCQWs) \cite{asymptotic1}, in the text we use quantum walks (QWs) to denote both. However, we conjecture that similar effects can be found in other quantum systems. Our motivation for analyzing transport properties of PCQWs is threefold. First, PCQWs constitute a modification of CQWs designed to describe walker's motion in dynamically changing medium. As these changes are not controlled, the resulting open dynamics involves decoherence. Second, there is a general theory allowing to construct trapped states for Grover PCQWs on a broad class of graphs \cite{Mares2019}. Third, an analytical treatment of PCQWs provides also significant insight into transport properties of CQWs. Thus, throughout the whole paper we investigate transport properties for PCQWs analytically and subsequently obtained favourable cases are examined for CQWs numerically.

\begin{figure}
    \centering
    \includegraphics[width=180 pt]{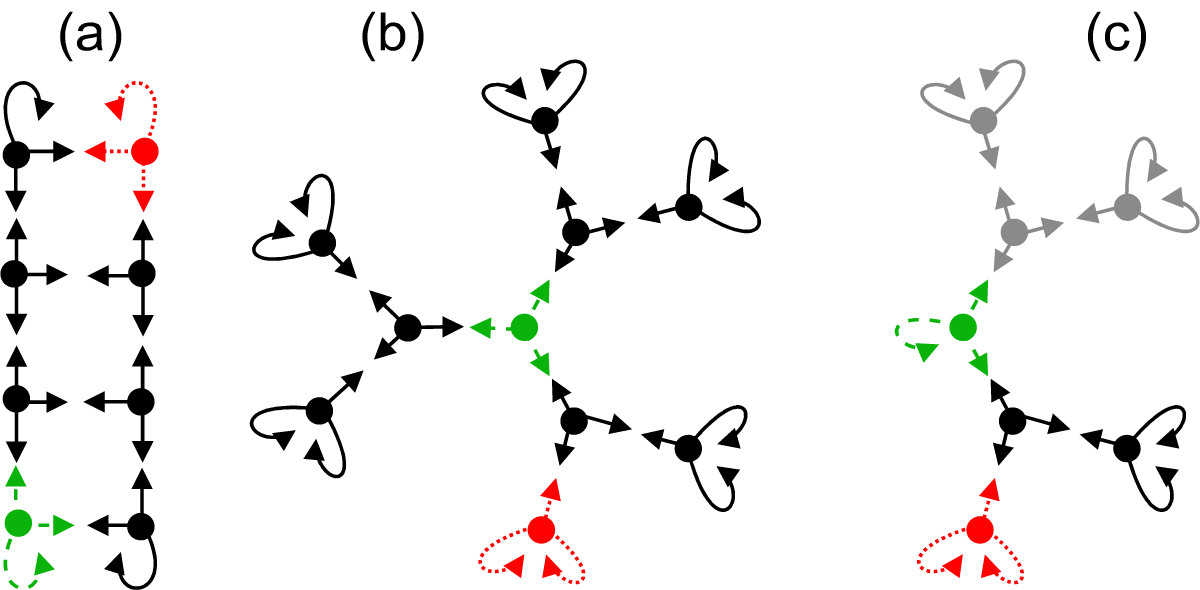}
    \caption{The ladder graph of the length $L=3$ (a), the Cayley tree of the order $k=2$ (b) and the "reduced"~Cayley tree with one or two branches missing (c) (removed and gray respectively). The vertices correspond to possible positions of the walker and the edges to the base computational states. The initial subspace is depicted by green dashed lines and the sink subspace by red dotted lines.}
    \label{fig:structures}
\end{figure}

As we intend to analyze transport properties of QWs on complex graphs, we employ a well suited definition of QWs provided in \cite{Mares2019}. Here, vertices $V$ of a directed state graph $G(V,E=E_p \cup E_l)$ represent all possible walker's positions and its edges $E$ encode directions which walker will follow in the next step of the evolution. We distinguish pairs of arcs $e,\bar{e} \in E_p$ with mutually opposite directions representing traversable links and additional self-loops $E_l$. Examples of state graphs employed in this work are depicted in FIG. \ref{fig:structures}. Every directed edge $e=(v_1,v_2) \in E$ (with $v_i \in V$) corresponds to a base state $\ket{e}$ of the walker's Hilbert space $\sH = \mathrm{span}\{\ket{e} | e \in E\}$ and represents the walker at position $v_1$ facing towards the vertex $v_2$. A general pure state of the walker reads
\begin{equation}
\ket{\psi} = \sum_{e \in E} \psi_{e} \ket{e}.
\label{general_pure_state}
\end{equation}
In this work we restrict ourselves to QWs on 3-regular graphs, i.e. each vertex has 3  outgoing and 3 incoming edges including eventual self-loops. Each triplet of edges originating in a common vertex $v$ defines the vertex subspace $\sH_v$, thus $\sH = \bigoplus_{v\in V} \sH_v$. The initial point (source) and sink of the walk are vertices of the graph $i$ and $s$ respectively with associated vertex subspaces $\sH_i$ and $\sH_s$.

In CQW, the walker's time evolution is realized in repeated steps, each consisting of three subsequent operations. First, the unitary reflecting ("flip-flop") shift operation
\begin{align*}
R = \sum_{e \in E_p} \ket{\bar{e}}\bra{e} + \sum_{e\in E_l} \ket{e}\bra{e},
\end{align*}
swaps amplitudes of arcs going in opposite directions and leaves the amplitudes of loops unchanged. Then, it is followed by unitary coin operation $C = \bigoplus_{v\in V} C_v$ which locally shuffles amplitudes in each individual 3-dimensional vertex subspace. We choose the extensively studied Grover matrix
\begin{align*}
C_v = G_3 =\frac{1}{3}\left[
\begin{array}{rrr}
 -1 & 2 & 2 \\
 2 & -1 & 2 \\
 2 & 2 & -1 \\
\end{array}
\right]
\end{align*}
for all vertices. The reflecting shift operator, well defined on all state graphs, is a natural choice, which in combination with the Grover coin establishes dynamics equipped with a rich structure of trapped states \cite{Inui2004}. Other choices of the shift operator can be readily implemented by an appropriate modification of the coin \cite{Mares2019}. Finally, at the end of each step, part of the walker's wave function, which has reached the sink, is effaced. The impact of the sink is represented by a non-unitary operation, projection onto the complement of the sink subspace $\Pi = I-\Pi_{\sH_s}$. Altogether, one step of CQW maps the walker's state density operator $\rho(t)$ at time $t$ to
\begin{align*}
\rho(t+1) = \Pi CR\rho(t)RC\Pi.
\end{align*}

PCQWs address quantum transport in a dynamically changing medium. In each step, some edges become  closed (or open again), simultaneously in both directions, not allowing the walker to pass between previously connected vertices. As this process is random and uncontrolled, each configuration of open edges $K \subset E_p$ appears with some probability $\pi_K$ and corresponds to a modified reflecting shift operator
\begin{align*}
R_K = \sum_{e \in K} \ket{\bar{e}}\bra{e} +\sum_{e \in E \setminus K} \ket{e}\bra{e}.
\end{align*}
One step of the walker's evolution incoherently incorporates all possible walker's paths and it is described by the superoperator $\Phi$
\begin{align*}
\rho(t+1) = \Phi(\rho(t)) = \sum_{K\subset E_p} \pi_K \Pi CR_K\rho(t)R_KC\Pi.
\end{align*}

QWs with sink are not trace-preserving processes and $\Tr(\rho(t))$ expresses the probability at time $t$ that the walker is still present at some vertices of the graph. Consequently, the total probability of the walker being absorbed in the sink
\begin{equation}
\label{def_efficiency}
q = \lim_{t \rightarrow +\infty} q(t)= 1 - \Tr\left(\lim_{t \rightarrow +\infty} \rho(t)  \right),
\end{equation}
defines the overall efficiency of the transport. It ranges from 0 to 1 and it is determined by the overlap between the walker's initial state and the subspace of so-called sr-trapped states ("sink resistant"), see \cite{Mares2019,assisted_transport}. These are trapped states which have no overlap with the sink subspace and thus survive the evolution in the presence of the sink. If their orthonormal basis is known, we can construct a projector $\Pi_T$ onto this subspace and write down (\ref{def_efficiency}) for any initial walker's state $\rho_{i}$ as $q=1-\Tr(\Pi_T \rho_{i})$. The detailed theory of trapped states for percolated reflecting Grover quantum walks on 3-regular graphs was developed in \cite{Mares2019}. This enables us to investigate various scenarios on different graphs, with different positions of the sink and very importantly for various initial states. We typically choose those for which the studied effects are the most pronounced and compare them with the average transport efficiency $\bar{q}$ (averaged over all states from the initial subspace) given by $\bar{q}= 1- \Tr(\Pi_T \overline{\rho})$ with $\overline{\rho}$ denoting the maximally mixed state on the subspace $\sH_i$.
\begin{figure}[t]
    \centering
    \includegraphics[width=180 pt]{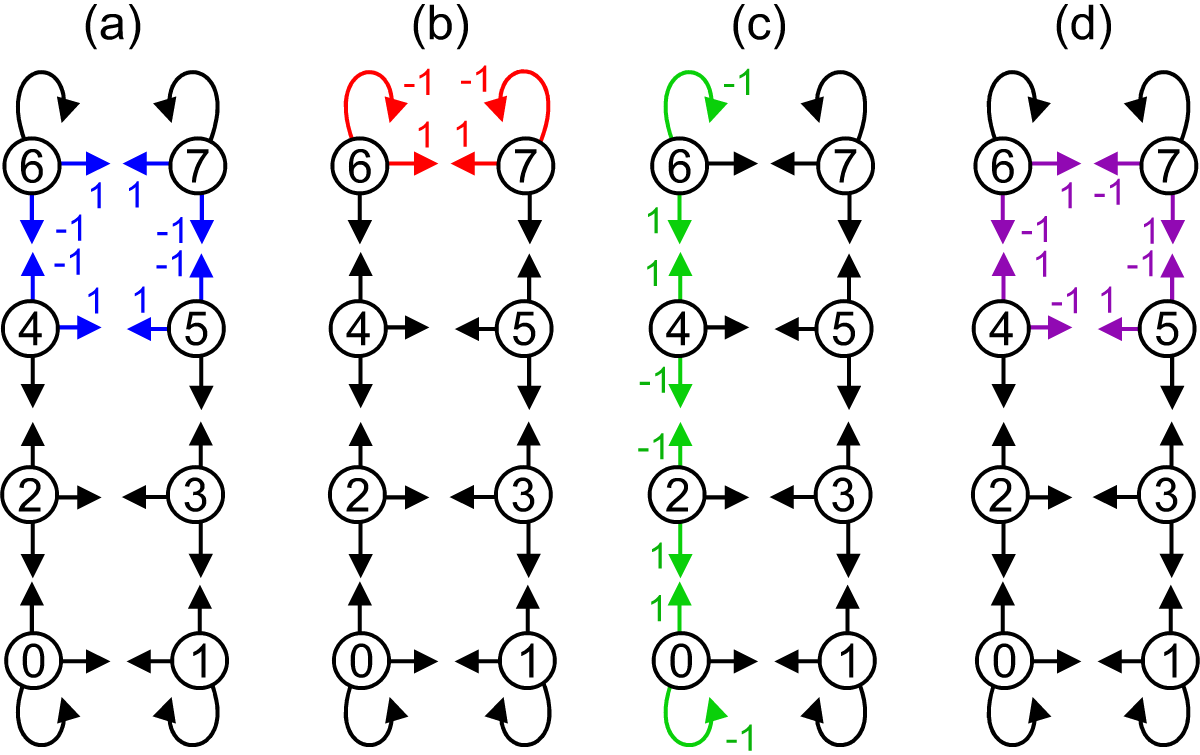}
    \caption{Linearly independent trapped states of the Grover CQWs and PCQWs on the ladder graph with $L=3$. There are $3$ "face-cycle"~states (a), two "short path"~states (b) and one "connecting path"~ (c). For the non-percolated CQW we have $3$ additional trapped states (d). Weights of graphs are nonzero coefficients of corresponding trapped states.}
    \label{fig:trapped_ladder_all}
\end{figure}

Let us first investigate how the distance between the initial and sink position affects transport efficiency. Our intuition tells us that longer distance implies less efficient transport. We show that quantum transport over longer distances may be more efficient. To illustrate the mechanism we choose a simple graph structure, yet exhibiting this effect. 
We first focus on the ladder graph, see FIG. \ref{fig:structures}(a), characterized by its length $L$, i.e. the number of square faces. The initial position and associated states are marked by green dashed lines and the sink with associated states by red dotted lines. The associated Hilbert space has dimension $6(L+1)$. All trapped states of Grover PCQWs on 3-regular graphs are associated with eigenvalue $-1$ and in this particular case their subspace has dimension $L+3$. Their basis is formed by $L$ "face-cycle"~states, two "short path"~states and one "connecting path"~state, whose examples are depicted in FIG. \ref{fig:trapped_ladder_all} for $L=3$. Weights denote nonzero non-normalized coefficients $\psi_e$ in (\ref{general_pure_state}) of corresponding trapped states, e.g. the trapped state in FIG. \ref{fig:trapped_ladder_all}(b) is $\ket{\psi_{(b)}} = -\ket{(6,6)}+\ket{(6,7)}+\ket{(7,6)}-\ket{(7,7)}$. Note that all trapped states are presented before their mutual orthonormalization, a necessary step to construct the projector $\Pi_T$.

\begin{figure}
    \centering
    \includegraphics[width=200 pt]{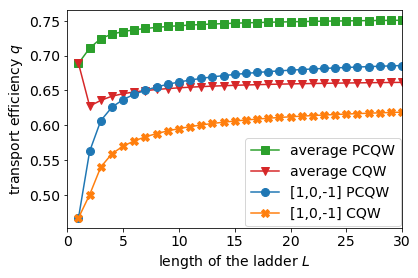}
    \caption{The source-to-sink transport efficiency of Grover CQW and Grover PCQW for different lengths of the ladder: the transport with the initial state $\ket{\psi_0}$ for percolated walk (blue circles) and for non-percolated walk (orange crosses), and the averaged transport efficiency  for percolated walk (green squares) and non-percolated walk (red triangles).}
    \label{fig:transport_ladder}
\end{figure}

Now, increasing the length of the ladder initiates two competing mechanisms. First, the number of trapped states grows resulting possibly in lower transport efficiency. On the other hand, as the "connecting path"~trapped state is stretched along the whole ladder, its individual coefficients $\psi_e$ are diminished with increasing $L$ due to normalization. It suggests that despite the first mechanism walker's initial states having a significant overlap with this trapped state may eventually show more efficient transport for longer ladders. Indeed, as the plot in FIG. \ref{fig:transport_ladder} shows transport efficiency surprisingly increases with the distance separating the source and the sink. We observe considerable efficiency increase of over 20 \% for the initial state $\ket{\psi_0}=  \frac{1}{\sqrt{2}}(\ket{(0,0)} - \ket{(0,2)}) \equiv \frac{1}{\sqrt{2}}[1,0,-1]$ (further in the text we use this notation for coefficients in the initial subspace), which maximizes the overlap with the "connecting path"~trapped state. The effect is present and significant even if we analyze the average transport efficiency $\bar{q}$.

Naturally, one could argue that the revealed transport effect is a direct consequence of the percolation. As plots in FIG. \ref{fig:transport_ladder} indicate, this is not true. According to \cite{Mares2019}, any trapped state of PCQWs is a trapped state of the corresponding CQW. Thus the argument with the "connecting path"~trapped state holds for the Grover CQW on ladder too. However, the Grover CQW on ladder is equipped with additional $L$ independent trapped states depicted in FIG. \ref{fig:trapped_ladder_all} \cite{note1}. This implies two consequences. First, percolation destroys some trapped states of CQWs and therefore enhances the overall source-to-sink transport. It is an example of the well known environment-assisted transport enhancement. Second, the number of trapped states of the CQW grows faster compared with the case of the PCQW and thus the effect of higher transport efficiency for longer ladders is less pronounced in this case.

The second family of graphs, whose transport properties we investigate, are 3-regular Cayley trees, i.e. each of their non-leaf vertex has 3 neighbors and all leaves have two loops and the same distance to the root \cite{Ostilli}. Order of the Cayley tree is the number of its shells $k$, sets of vertices with the same distance to the root (see example in FIG. \ref{fig:structures}(b) with $k=2$). Employing Cayley trees we can study two different changes of geometry and its influence on the transport efficiency from tree root to one of its leaves.

\begin{figure}
    \centering
    \includegraphics[width=200 pt]{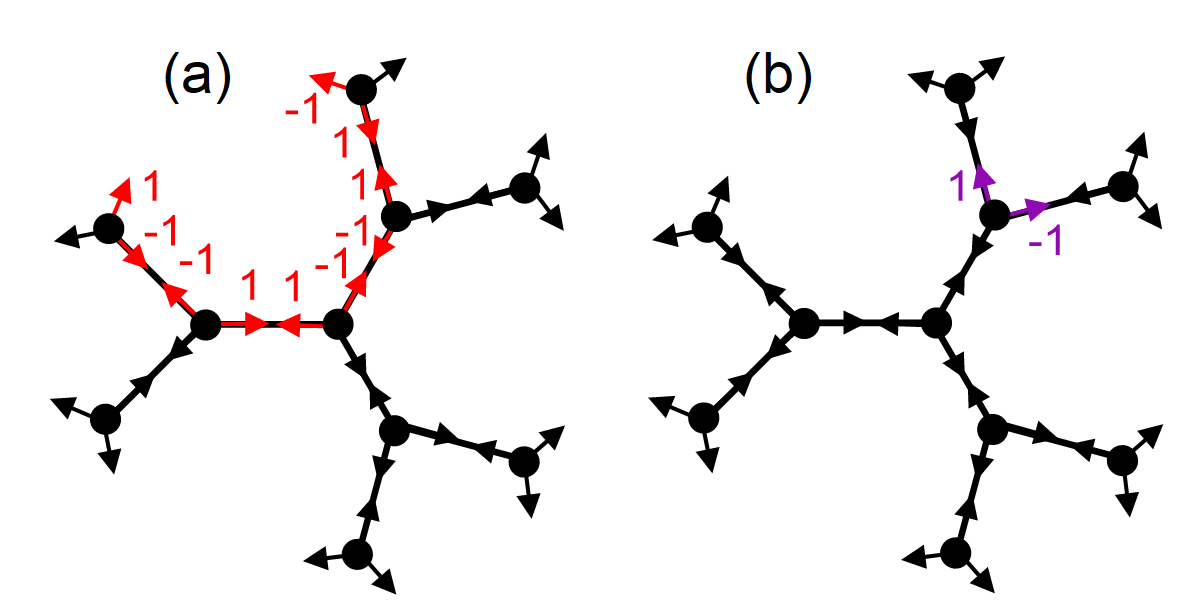}
    \caption{An example of "path"~state of Grover PCQW on the Cayley tree (a). An example of additional trapped state for the Grover CQW on the Cayley tree (b).}
    \label{fig:trapped_snowflake}
\end{figure}

We first study extension of the Cayley tree by increasing its order. As the number of edges and so the dimension of the Hilbert space $\dim(\sH_k)=3(1+3(2^k-1))$ grows exponentially, we are facing a scenario very different from the linear structure of the ladder. All trapped states of the Grover PCQW correspond to eigenvalue $-1$ and their dimension is $3\cdot 2^k-1$. Embedding the Cayley tree into a plane without crossing its edges, we can choose their basis as "path"~states represented by shortest paths between neighboring loops, see example depicted in FIG. \ref{fig:trapped_snowflake} (a).
Altogether, these are $3\cdot 2^k$ linearly dependent, non-normalized, ”path” states. Orthogonalization can be done so that only two resulting sr-trapped states, denoted as $\ket{T_1}$ and $\ket{T_2}$, overlap with the initial subspace $\sH_i$. For details see Supplementary material.

\begin{figure}
    \centering
    \includegraphics[width=200 pt]{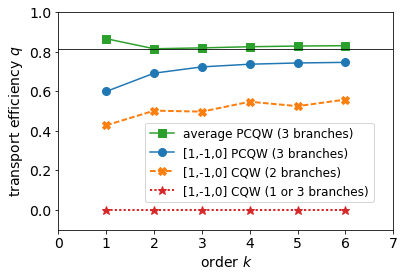}
    \caption{The transport efficiency of the Grover PCQW on the Cayley tree (blue circles) and the Grover CQW on the reduced Cayley tree with two branches (orange crosses) and with one branch (green diamonds) for the initial state $\ket{\psi_1}$, the averaged transport efficiency (green squares) for Grover PCQW on the Cayley tree of different orders $k$.}
    \label{fig:transport_snowflake_percolated}
\end{figure}

As we enlarge the Cayley tree, exponentially growing number of new trapped states increases coefficients of  $\ket{T_1}$ and $\ket{T_2}$ in $\sH_i$. Simultaneously, these coefficients are diminished by normalization thanks to additional non-zero amplitudes out of $\sH_i$. The initial decrease in average transport efficiency of PCQW in FIG. \ref{fig:transport_snowflake_percolated} confirms that compared to the transport on the ladder it is harder to find an initial state exhibiting higher efficiency by extension of the Cayley tree. However, we present such state. The maximal obtained increase of transport efficiency for Grover PCQW on the Cayley tree shown in FIG. \ref{fig:transport_snowflake_percolated} is attained for the initial state $\ket{\psi_1}=\frac{1}{\sqrt{2}}[1,-1,0]$ (the zero coefficient is assigned to the sink-branch edge), which maximizes the overlap with the state $\ket{T_1}$. The transport efficiency for this state is minimal in the PCQW and is given by the formula
\begin{align}
\label{atp_analytical}
q(k)=1-\frac{2^k}{2^{k+2}-3}.
\end{align}
Proofs of these statements are provided in Supplementary material. Moreover, this initial state shows zero transport efficiency of the CQW for all orders of the Cayley tree. This is another example of a strong environment-assisted transport enhancement boosting transport efficiency from 0\% to 75\% for $k\rightarrow\infty$. 

\begin{figure}[t!]
    \centering
    \includegraphics[width=200 pt]{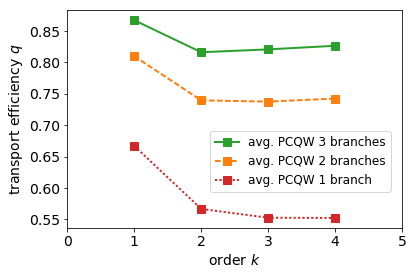}
    \caption{The averaged transport efficiency of Grover PCQW on the Cayley tree and its reduced modifications for different orders $k$.}
    \label{fig:snowflake_normal_vs_disabled_averaged}
\end{figure}

Finally, the geometry of Cayley trees offers to investigate how the transport efficiency is affected if we cut main branches without the sink and replace them by loops, see incurred reduced Cayley trees depicted in FIG. \ref{fig:structures}(c). Shells and order of reduced Cayley trees are defined analogously. Despite obvious expectation, we find that the removal of dead-end branches may actually decrease the averaged transport efficiency of PCQW as it is shown in FIG. \ref{fig:snowflake_normal_vs_disabled_averaged}. The reason for this behaviour may be found in the decisive role of the trapped states $\ket{T_1}$ and $\ket{T_2}$. Cutting branches solely truncates corresponding parts of both states outside of $\sH_i$ which due to normalization amplifies their coefficients and consequently increases walker's trapping.

Turning attention to the CQW on Cayley trees, we emphasize that the walker initiated in the state $\ket{\psi_1}$ is fully trapped. Cutting one of the main branches may only improve transport in this case, as exemplified in FIG. \ref{fig:transport_snowflake_percolated}. But if we cut both main branches, the efficiency drops back to zero. We observe again that adjunction of branches without sink may result in higher transport efficiency. Moreover, in this particular case transport efficiency is improving with the extension of the reduced Cayley tree.

In conclusion, we have demonstrated that for the Grover CQWs and PCQWs on ladder and Cayley trees the transport efficiency may be enhanced by increasing distance between source and targeted sink or by adding redundant dead-end branches to the actual graph. Since all observed quantum effects result from the interplay between the particular set of trapped states and the initial state of the QW, we expect that these phenomena might be identified also in other quantum systems. Using a complete description of trapped states for PCQWs we explained these effects and found initial states well exhibiting these effects also for CQWs. We stress that especially the Grover CQWs on Cayley trees are equipped with very rich structures of trapped states, not well explored yet. Consequently, their transport efficiency may show even more complex behavior, as we extend the graph or add redundant branches. Investigations of these transport phenomena go beyond the scope of this paper and are left for future studies.

All authors acknowledge the financial support from RVO14000 and "Centre for Advanced Applied Sciences", Registry No. CZ.02.1.01/0.0/0.0/16\_019/0000778, supported by the Operational Programme Research, Development and Education, co-financed by the European Structural and Investment Funds and the state budget of the Czech Republic. JN, JM and IJ are supported by the Czech Science foundation (GA\v CR) project number 16-09824S and by Grant Agency of the Czech Technical University in Prague, grant No. SGS16/241/OHK4/3T/14.  M\v{S} and IJ are partially supported from GA\v{C}R 17-00844S.

\section{Supplementary material}
In this part we present a basis of trapped states for the Grover PCQWs on Cayley trees followed by the proof of the statement that the minimal transport efficiency is achieved for the state $\ket{\psi_1}=\frac{1}{\sqrt{2}}[1,-1,0]$ and it is given by the formula (3). 

A general recipe for the construction of a (non orthogonal) basis of the trapped-states subspace is given in \cite{Mares2019}. It originates from two underlying rules: the two elements for arcs going in the opposite directions must be equal, and the sum of elements for edges originating in one vertex has to be zero.  It is proven therein that the complete basis of $3\cdot 2^k -1$ independent trapped states and $3\cdot 2^k -3$ sr-trapped states after removing those overlapping with the sink can be obtained from "path" states (see FIG. 4(a)). 
Each path state, connecting always two neighboring loops, begins and terminates by a loop and its elements are ones and minus ones oscillating along a path while following the two rules above. The whole basis is chosen in such a way, that only two of its elements have nonzero overlaps with the initial subspace $\sH_i$.

However, this basis of sr-trapped states is non-orthogonal and for the calculation of the transport efficiency we need its (in our case partial) orthogonalization. For Cayley trees it can be achieved in two steps. First, we construct a new basis of sr-trapped states. By a one-to-one correspondence we assign to each sr-trapped path state an element of newly constructed basis. Starting with the given path state we add to its support all branches which are attached in the Cayley graph to its vertices except the middle point (the vertex which splits the path state into two anti-symmetric branches). If an added branch contains the sink, we add this branch without the sink vertex and all its outgoing and incoming edges and call this element of the new basis \emph{sink-affected}.
This defines support of the constructed element of the new basis. Coefficients of both edges attached to its middle point are set to have values $2^{k-m}$ and $-2^{k-m}$, where $m$ is distance of the middle point and the root of the Cayley tree. Now moving from the middle point towards its leaves, the state coefficients are halved after every branching of the state support and their signs are flipped after passing any vertex. 

\begin{figure}
    \centering
    \includegraphics[width=180 pt]{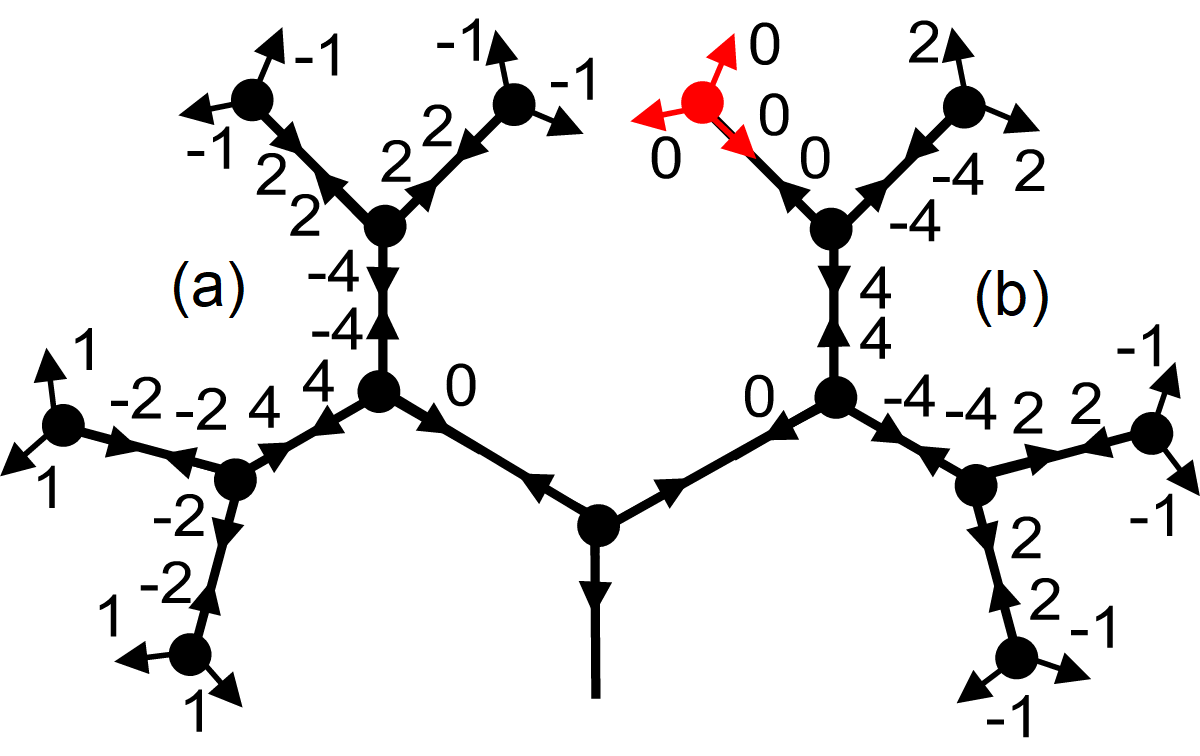}
    \caption{Elements of constructed basis of sr-trapped states. A sink-not-affected vector (a) and sink-affected vector (b). The sink is marked by red color. These are two distinct states shown in one figure.}
    \label{fig:example_orthogonal}
\end{figure}

The obtained new basis of sr-trapped states contains two sets of elements. The first set contains $k-1$ mutually non-orthogonal sink-affected elements. Let us denote them with their increasing supports as $\ket{u_1}, \ket{u_2}, \ldots, \ket{u_{k-1}}$. An example of $\ket{u_1}$ for $k=3$ is depicted in FIG. \ref{fig:example_orthogonal}(b). The second set consists of the remaining sink-not-affected elements which are mutually orthogonal. An example is the element shown in FIG. \ref{fig:example_orthogonal}(a), which is constructed from the path state 4(a). 
Another example of such states are path states connecting both loops of one leaf. Their orthogonality directly follows from the fact that each of them has two anti-symmetrical branches split by the middle point, i.e. these two branches have same magnitudes of coefficients with opposite signs. Therefore, different sink-not-affected elements have either no common support or they share a support with corresponding anti-symmetric and symmetric coefficients.  For the same reason elements from both sets are mutually orthogonal except the two whose support contains the root of the Cayley tree (states $\ket{u_{k-1}}$ and $\ket{t_1}$, see below).

\begin{figure}
    \centering
    \includegraphics[width=110 pt]{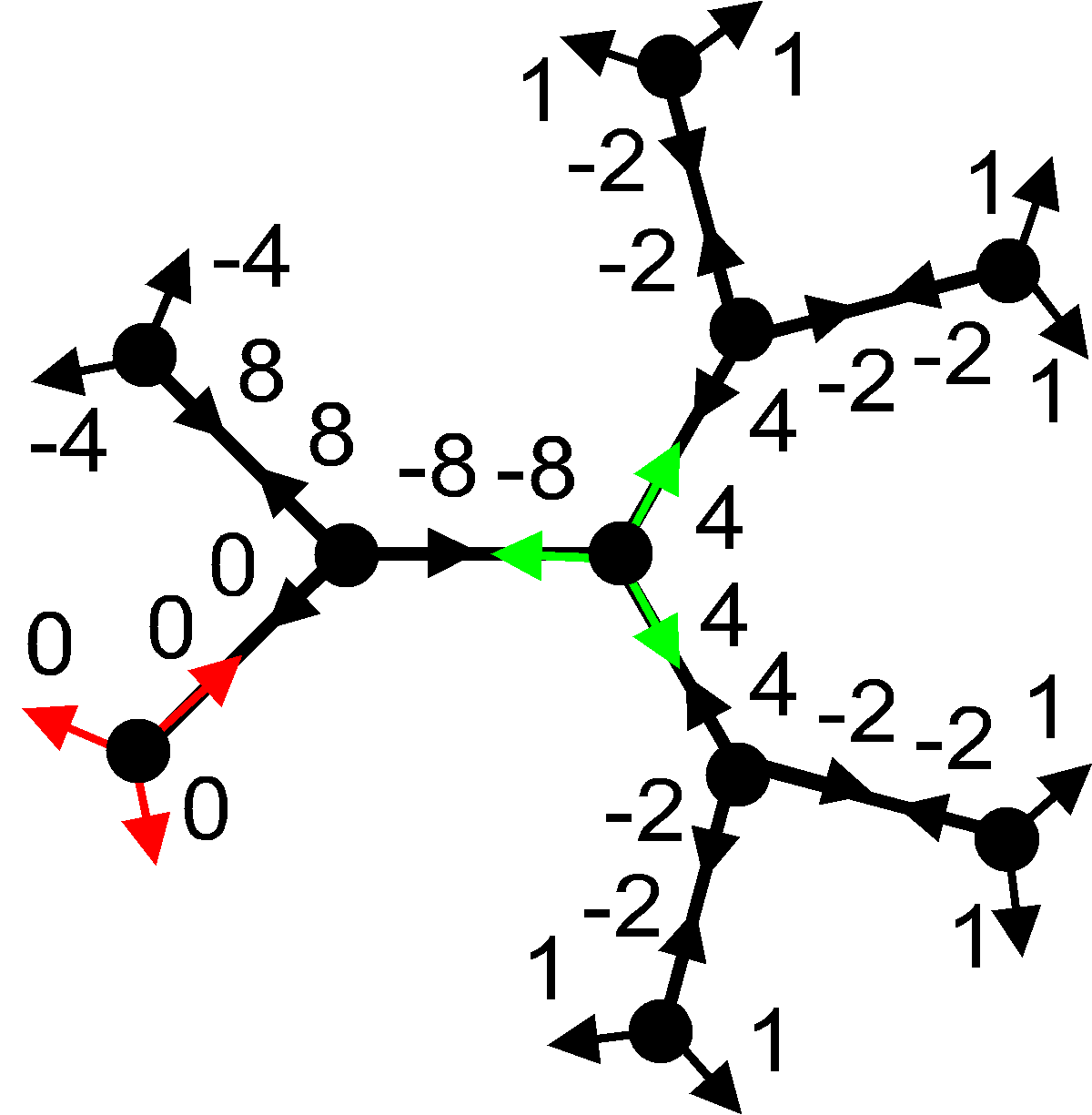}
    \caption{Example of a sr-trapped state $\ket{\tilde{t}_2}$ for the order $k=2$.}
    \label{fig:t2_state}
\end{figure}

At this point, let us recall that the transport efficiency may be expressed as $q=1-\Tr\{\Pi_T \rho_i\}$. Thus, in order to calculate the transport efficiency we actually need only those elements from orthogonal basis of sr-trapped states which have a nonzero overlap with the initial subspace $\sH_i$. We find them in the second step of our procedure. In our new basis there are only two vectors overlapping with the initial subspace for $k>1$ (the case with $k=1$ is trivial and we skip it). The first one, denoted as $\ket{t_1}$, belongs to the set of sink-not-affected elements and has a nonzero overlap solely with the element $\ket{u_{k-1}}$, which is the second vector having a nonzero overlap with the initial subspace. Due to our construction, these vectors satisfy the relation
$$
\braket{t_1,t_1}=-2\braket{t_1,u_{k-1}}
$$
and this allows us to replace $\ket{u_{k-1}}$ in our basis by the vector $\ket{\tilde{t}_2} = 2\ket{u_{k-1}} + \ket{t_1}$. This makes the vector $\ket{t_1}$ orthogonal to all other elements of the basis. On the other hand, the vector $\ket{\tilde{t}_2}$, whose example for $k=2$ can be seen in FIG. \ref{fig:t2_state}, is not orthogonal to the other sink-affected elements $\ket{u_1},\ket{u_2}, \ldots \ket{u_{k-2}}$ yet. In principle, we could employ the Gram-Schmidt orthogonalization process to replace the vector $\ket{\tilde{t}_2}$ in our basis by the new vector $\ket{t_2}$ orthogonal to all other elements of the basis. However, we do not need the explicit form of the state $\ket{t_2}$. It is sufficient to show that its norm satisfies 
\begin{equation}
\label{eq_norm_inequality}
||t_2||^2 > 3||t_1||^2.    
\end{equation}
This can be done in the following way. In general, the vector $\ket{t_2}$ can be written in the form 
$$
\ket{t_2}= 2\ket{u_{k-1}} + \ket{t_1} + \sum_{i=1}^{k-2} \alpha_i \ket{u_i} = \sum_{e \in E} \phi_e \ket{e}.
$$
Since each of the states $\ket{u_i}$ except $\ket{u_{k-1}}$ has zero overlap with the initial subspace, coefficients $\phi_e$ associated with base states in the initial subspace of the state $\ket{t_2}$ are still $2^k,2^k,-2^{k+1}$, where the last element corresponds to the edge belonging to the branch with the sink. Moreover, the rule that the sum of coefficients at each vertex is zero must hold. Among all vectors, which follow these two rules, the minimal norm is achieved for the vector whose coefficients are halved after each branching (going from the root towards leaves). Apparently, this vector $\ket{w}$ is a result of the same construction as the vector $\ket{t_2}$ but now without the presence of the sink. In particular, $\ket{u_{k-1}}$ is a sink-affected vector, i.e. its construction from the path state was influenced by the presence of the sink. We construct new vector $\ket{\tilde{w}}$ from the same path state but now regardless of the sink. As both vectors $\ket{\tilde{w}}$ and $\ket{t_1}$ originate from path states going through the root of the Cayley tree, they follow equations
$$
\braket{t_1,t_1}=\braket{\tilde{w},\tilde{w}}=-2\braket{t_1,\tilde{w}}.
$$
Therefore, the vector $\ket{w}=2\ket{\tilde{w}}+\ket{t_1}$ fulfills $||w||^2=3||t_1||^2$ and as $||t_2|| > ||w||$ holds, relation (\ref{eq_norm_inequality}) is proved.


Equipped with partially orthogonal basis of sr-trapped states we can investigate properties of the transport efficiency. For the order $k$ the state $\ket{t_1}$ has elements $2^k$, $-2^k$ and $0$ in the initial vertex and the state $\ket{t_2}$ has elements $2^k$ on the non-sink branches and $-2^{k+1}$ on the sink branch. We denote their normalized forms as $\ket{T_1}\equiv \frac{\ket{t_1}}{||t_1||}$ and $\ket{T_2}\equiv \frac{\ket{t_2}}{||t_2||}$. The restriction of the vector $\ket{T_1}$ (resp. $\ket{T_2}$) to the initial subspace is $\ket{\tau_1} = \frac{2^k[1,-1,0]}{||t_1||}$ (resp. $\ket{\tau_2} = \frac{2^k[1,1,-2]}{||t_2||}$). Taking into account the inequality (\ref{eq_norm_inequality}) we straightforwardly obtain $||\tau_2|| < ||\tau_1||$. Let us also note that $\braket{\tau_1 | \tau_2} = 0$.

Considering an arbitrary initial walker's state represented by a unit vector $\ket{X}$ with its restriction to the initial subspace $\ket{x}$, the probability of trapping may be expressed and bounded as 
\begin{align}
P(k,\ket{x}) &= \Tr\left\{\Pi_T \ket{X}\bra{X}\right\} = |\braket{T_1,X}|^2+|\braket{T_2,X}|^2  \nonumber \\ &= |\braket{\tau_1,x}|^2+|\braket{\tau_2,x}|^2 \nonumber \\
&=||\tau_1||^2\left| \braket{\frac{\tau_1}{||\tau_1||}, x } \right|^2 + ||\tau_2||^2\left| \braket{\frac{\tau_2}{||\tau_2||}, x } \right|^2  \nonumber \\ &\leq  ||\tau_1||^2\left( \left| \braket{\frac{\tau_1}{||\tau_1||}, x } \right|^2 + \left| \braket{\frac{\tau_2}{||\tau_2||}, x } \right|^2 \right) \nonumber \\
&\leq  ||\tau_1||^2,
\end{align}
where we use the Bessel's inequality in the last inequality. Since the maximal trapping is achieved for the state $\ket{\psi_1}=\frac{1}{\sqrt{2}}[1,-1,0] = \frac{\ket{\tau_1}}{||\tau_1||}$ we conclude that there is no initial state with a lower transport efficiency.

\begin{figure}
    \centering
    \includegraphics[width=200 pt]{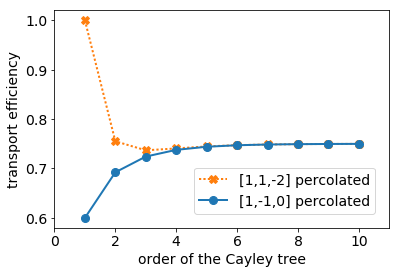}
    \caption{Transport efficiency of PCQW on the Cayley tree depicted for the initial state $\ket{\psi_1}$ (blue full, circles) and $\ket{\psi_2}$ (orange dotted, crosses).}
    \label{fig:nemonotonnost_t2}
\end{figure}

The symmetrical form of the state $\ket{T_1}$ allows for determination of the normalization constant
\begin{align*}
||t_1(k)||^2=2^{k+1}+\sum_{i=0}^{k-1} 2^{2+i}(2^{k-i})^2 = 2^{k+1}(2^{k+2}-3).
\end{align*}
Hence, the transport efficiency for the initial state $\ket{\psi_1}$ is given as
\begin{align*}
q(k)=1-|\braket{\psi_1 | T_1}|^2 = 1- \frac{(2\cdot 2^{k})^2}{2\cdot ||t_1(k)||^2}=1-\frac{2^k}{2^{k+2}-3}.
\end{align*}
Apparently, the transport efficiency grows with increasing order of the Cayley tree. Writing the trapping probability as
\begin{equation*}
P(k,\ket{x}) = \frac{4^k}{||t_1\frac{\ket{\tau_2}}{||\tau_2||} = (k)||^2}\left|\braket{\psi}\right|^2+\frac{4^k}{||t_2(k)||^2}\left|\braket{\psi_2,x}\right|^2,
\end{equation*}
where we define $\ket{\psi_2} = \frac{\ket{\tau_2}}{||\tau_2||} = \frac{1}{\sqrt{6}}[1,1,-2]$ ,
we find out that the maximal increase of the transport efficiency between two orders $k_1$ and $k_2$ is always achieved either for the walker's initial state $\ket{\psi_1}$ or $\ket{\psi_2}$. A comparison of their transport efficiencies depicted in Fig. \ref{fig:nemonotonnost_t2} shows that the effect is the most pronounced for the initial state $\ket{\psi_1}$.
Finally, we point out that the opposite task is simple. The initial state $\frac{1}{\sqrt{3}}[1,1,1]$ is orthogonal to both $\ket{\psi_1}$ and $\ket{\psi_2}$ and so it results in complete transport regardless of the actual value $k$.


\begin{thebibliography}{99}
\bibitem{note1}
A complete theory of trapped states for CQWs is not developed. However, it is straightforward to check that all these additional $L$ states are trapped. Numerical simulations performed for a sufficiently large range of the length $L$, confirm that we always construct projector $\Pi_T$ from a complete basis of trapped states for the Grover CQW on the ladder.
\end{thebibliography}
\end{document}